\documentclass[epjST]{svjour}
\usepackage{graphics}
\usepackage{url}

\usepackage{graphicx,type1cm,eso-pic,color}

\begin{document}
\title{A planetary nervous system for social mining and collective awareness}
\author{Fosca Giannotti\inst{1}\fnmsep\thanks{\email{fosca.giannotti@isti.cnr.it}}
Dino Pedreschi\inst{2}\fnmsep\thanks{\email{pedre@di.unipi.it}}
Alex (Sandy) Pentland\inst{3}\fnmsep\thanks{\email{pentland@mit.edu}}
Paul Lukowicz\inst{4}\fnmsep\thanks{\email{paul.lukowicz@dfki.de}}
Donald Kossmann\inst{5}\fnmsep\thanks{\email{donaldk@inf.ethz.ch}}
James Crowley\inst{6}\fnmsep\thanks{\email{james.crowley@inrialpes.fr}}
Dirk Helbing\inst{5}\fnmsep\thanks{\email{dirk.helbing@gess.ethz.ch}}
}

\institute{ISTI-CNR, National Research Council, Pisa, Italy \and
University of Pisa, Italy \and
MIT, Massachusetts, Boston, USA \and
DFKI, Kaiserslautern, Germany \and
ETH Zurich, Switzerland \and
INRIA Rhone-Alpes, France}

\abstract{
We present a research roadmap of a {\em Planetary Nervous System} (PNS), capable of sensing and mining the digital breadcrumbs of human activities and unveiling the knowledge hidden in the big data for addressing the big questions about social complexity. We envision the PNS as a globally distributed, self-organizing, techno-social system for answering analytical questions about the status of world-wide society, based on three pillars: \emph{social sensing}, \emph{social mining} and the idea of \emph{trust networks} and \emph{privacy-aware social mining}. We discuss the ingredients of a science and a technology necessary to build the PNS upon the three mentioned pillars, beyond the limitations of their respective state-of-art.
{\em Social sensing} is aimed at developing better methods for harvesting the big data from the techno-social ecosystem and make them available for mining, learning and analysis at a properly high abstraction level.
{\em Social mining} is the problem of discovering patterns and models of human behaviour from the sensed data across the various social dimensions by data mining, machine learning and social network analysis.
{\em Trusted networks} and {\em privacy-aware social mining} is aimed at creating a new deal around the questions of \emph{privacy} and \emph{data ownership} empowering individual persons with full awareness and control on own personal data, so that users may allow access and use of their data for their own good and the common good.
The PNS will provide a goal-oriented knowledge discovery framework, made of technology and \emph{people}, able to configure itself to the aim of answering questions about the pulse of global society. Given an analytical request, the PNS activates a process composed by a variety of interconnected tasks exploiting the social sensing and mining methods within the transparent ecosystem provided by the trusted network. The PNS we foresee is the \emph{key tool for individual and collective awareness for the knowledge society}. We need such a tool for everyone to become fully aware of how powerful is the knowledge of our society we can achieve by leveraging our wisdom as a crowd, and how important is that everybody participates both as a consumer and as a producer of the social knowledge, for it to become a trustable, accessible, safe and useful public good.
} 
\maketitle
\section{Our Visionary Approach}
\label{intro}
One of the most pressing, and fascinating, challenges of our time is understanding the complexity of the global interconnected society we inhabit. This connectedness reveals in many phenomena: in the rapid growth of the Internet and the Web, in the ease with which global communication and trade now takes place, and in the ability of news and information as well as epidemics, trends, financial crises and social unrest to spread around the world with surprising speed and intensity.

Ours is also a time of opportunity to observe and measure how our society intimately works: the big data originating from the digital breadcrumbs of human activities, sensed as a by-product of the ICT systems that we use every day, promise to let us scrutinize the ground truth of individual and collective behaviour at an unprecedented detail in real time.

\begin{figure}\centering
\resizebox{0.75\columnwidth}{!}{%
\includegraphics{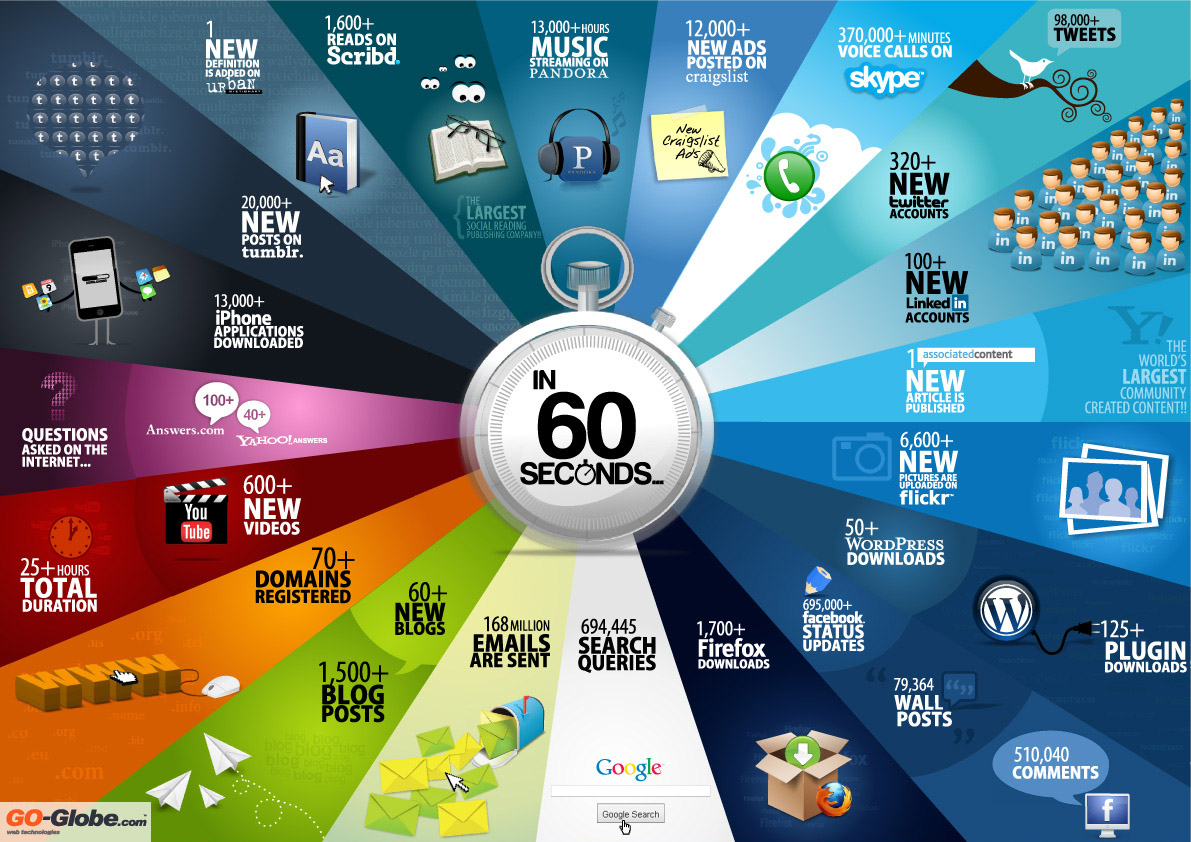} }
\caption{Things That Happen On Internet Every Sixty Seconds. By Shanghai Web Designers, http://www.go-gulf.com/60seconds.jpg.}
\label{userGeneratedContent}       
\end{figure}

Multiple dimensions of our social life have big data ``proxies" nowadays:
\begin{itemize}
\item our desires, opinions and sentiments leave their traces in our web pages and blogs, in the social media we participate in, in the query logs of the search engines we use, in the tweets we send and receive;
\item our relationships and social connections leave their traces in the network of our phone or email contacts, in the friendship links of our favourite social networking site;
\item our shopping patterns and lifestyles leave their traces in the transaction records of our purchases;
\item our movements leave their traces in the records of our mobile phone calls, in the GPS tracks of our on-board navigation system.
\end{itemize}
On an average day, people send around 47 billion email (spam excluded), post 100 million tweets and share 30 million pieces of contents on Facebook; Figure \ref{userGeneratedContent} depicts the astounding amount of user generated data produced every single minute. Other social-human data include financial, health and institutional data.
The fidelity of our digital breadcrumbs gets higher and higher, hand in hand with the sophistication of the sensors we interact with and carry with us, notably our smart phones. This increasing wealth of data is a promise of novel means for disentangling the societal complexity and face the challenges that the FuturICT project has set forth \cite{N1,N14,N7,N8}. It is clear how such challenges, such as the idea of a smart, self-organizing city, require high-level analytics, modeling and reasoning across all the social dimensions \cite{N16}. Authoritative sources such as the World Economic Forum \cite{Pentland2011} and The Economist \cite{Economist2010} advocate that we are just at the beginning of the data revolution era, which will impact profoundly all aspects of society – government, business, science, and entertainment. A frequently quoted sentence of Maglena Kuneva, former European Commissioner of Consumer Protection, maintains that ``personal data is the new oil of the Internet and the new currency of the digital world''.

In practice, however, the data revolution is in its infancy, and there is still a big gap from the big data to the big picture, i.e., from the opportunities to the challenges. The reasons behind the gap are the following three.
\begin{itemize}
\item Sensed data are fragmented, low-level and poor. They reside in diverse databases and repositories, often inaccessible for proprietary and legal constraints, and have limited power to portray different social dimensions concurrently.  Sensed data do not speak the language of human activities, but expose the raw details of the measurements allowed by the ICT infrastructure that generates them, so that it is difficult to respond to questions like: what is the activity associated with a person's geo-location recorded in mobile phone data? What is the purpose of a specific web query recorded in a search engine logs? The big size of data not always overcome semantic deficiency when modelling complex phenomena.
\item Analytics is too fragmented, low-level and poor, also because data mining models and patterns do not speak the language of human activities. Although data mining from traditional databases is a mature technology, we do not have yet a data mining for human activities, with their networked multi-dimensional nature and semantic richness, combined with big size and dynamicity. A similar consideration applies to network science and statistics: in the overall analytical process, the role of social semantics is confined to the preparation of data and to the interpretation of the obtained patterns and models, while limited is the ability of mining and statistical methods to take social semantics into account within appropriately high-level modelling. We still lack the ability to capture social diversity and complexity into patterns, models and indicators that narrate realistic stories, where the micro and macro levels of society are both considered and reconciled. In a nutshell, we lack holistic analysis, combining all social dimensions simultaneosly.
\item There are many regulatory, business and technological barriers to set the power of big data free for social mining, so that all individuals, businesses and institutions can safely access the knowledge opportunities. People are unaware of the power of their personal data, which are predated by service providers and the Internet companies and secluded in disparate public and private databases. Lack of trust and transparency makes people reluctant to share high-quality personal data. Current technologies and laws fall short of providing the legal and technical personal data ecosystem needed to support a well-functioning digital economy. As a result, there is limited privacy protection for the users whose data are passively captured by service providers, as well as little transparency on how such personal data are used. Moreover, the limited trust that we experience on-line, together with the absence of adequate privacy protection and incentives, prevent the engagement of people in effective campaigns for collecting social data.
\end{itemize}

\begin{figure}\centering
\resizebox{0.45\columnwidth}{!}{
\includegraphics{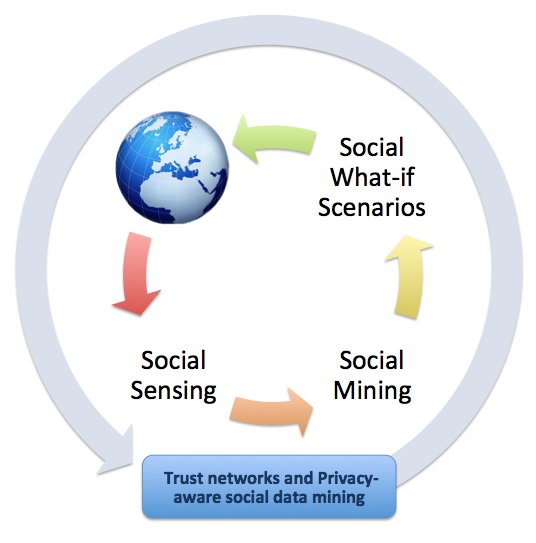} }
\caption{Social knowledge discovery process.}
\label{socialMiningProcess}       
\end{figure}

How can we remove the barriers for the rise of a knowledge-based society? The vision we advocate in this paper is that of a \emph{Planetary Nervous System}, PNS in short: a middle-layer technology for bridging the gap by making the knowledge and semantics hidden in the big data available for addressing the big questions about social complexity \cite{Pentland2012}. {\em We envision the PNS as a goal-oriented, globally distributed, self-organizing, techno-social system for answering analytical questions about the status of world-wide society, based on three pillars: \emph{social sensing}, \emph{social mining} and the idea of \emph{trust networks} and \emph{privacy-aware social mining}, that together form the social knowledge discovery process depicted in Figure \ref{socialMiningProcess}}.

\textbf{Social sensing} is aimed at developing better methods for harvesting the big data from the techno-social ecosystem and make such big data available for mining/analysis at a properly high abstraction level. Today’s sensed data fail to describe human activities at an adequate, multi-faceted semantic level, which makes it difficult to observe human behaviour. Social sensing aims at overcoming these limitations by novel semantic-based techniques that can map human behaviour, both (i) by automated annotation and enrichment methods that extract multi-dimensional social semantics from raw digital breadcrumbs, and (ii) by advanced participatory sensing systems, that explicitly involve people into the picture to consciously volunteer high quality personal data, leveraging crowd computing and serious games.

\textbf{Social mining} is the problem of discovering patterns and models of human behaviour from the sensed data across the various social dimensions: it is aimed at extracting multi-dimensional patterns and models from multi-dimensional data. To this aim, social mining needs novel concepts of multi-dimensional pattern discovery and machine learning, and of multi-dimensional social network analysis. The ultimate goal is to understand \emph{human behaviour} through:
\begin{itemize}
\item the discovery of \emph{individual social profiles},
\item the analysis of \emph{collective behaviors},
\item the analysis of \emph{spreading}, \emph{epidemics} and \emph{social contagion} over multi-dimensional networks,
\item the analysis of \emph{sentiment and opinion evolution}.
\end{itemize}
The key point of social mining is to combine the macro and micro laws of human interactions within a uniform multi-dimensional analytical framework, encompassing both the global models developed by complex network science and the local patterns discovered by data mining in the sensed data.

\textbf{Trusted networks} and \textbf{privacy-aware social mining} is aimed at creating a new deal around the questions of \emph{privacy} and \emph{data ownership}. A \emph{user-centric personal data ecosystem}, empowering individual persons by providing full awareness, control, exploitation and transparency on own personal data, can be the key to achieve trust in participatory social sensing, and boost the emergence of personal data as a new strategic asset for society. Advances in analysis of network data must be approached in tandem with understanding how to create value for the producers and owners of the data, while at the same time protecting the public good.  On the other hand, privacy-preserving techniques can be developed for social mining, which protect the personal space of the people represented in the data while supporting the safe dissemination of data and/or analyses (including data anonymization and homomorphic cryptography).

In this paper, we discuss the ingredients of a science and a technology necessary to build a PNS upon the three mentioned pillars; we concentrate on the roadmap of research needed to pursue this ambitious objective, beyond the limitations of the state-of-art in big data sensing, mining and analytics. However, our vision of the PNS goes beyond a toolkit for sensing and mining social big data. The PNS will be a \emph{techno-social ecosystem} offering a goal-oriented knowledge discovery framework, made of technology and \emph{people}, able to configure itself towards the aim of answering questions about the status of the global society. Given an analytical request, such as for instance the computation of multi-dimensional indices of societal well-being, the PNS activates a process composed by a variety of interconnected tasks exploiting the full repertoire of available methods for sensing (including both participatory and opportunistic sensing) and mining within the transparent ecosystem provided by the trusted network. We conclude the paper by briefly discussing how the PNS might support the entire social knowledge discovery process needed to create, monitor and forecast a novel generation of indicators of social well-being that, beyond the limitations of the GDP and other current mainstream indices, strive to encompass multi-dimensional socio-economic performance instead of purely economic performance, as well as micro-level measurements instead of purely macro-level aggregates \cite{N7,N8}.

At the end of the day, the PNS we foresee is the \emph{key tool for individual and collective awareness for the knowledge society} we inhabit. We need such a tool for everyone to become fully aware of two intertwined evidences: first, how powerful is the knowledge of our society we can achieve by leveraging our wisdom as a crowd, at all levels from local communities to global population, and second, how important is that everybody participates both as a consumer and as a producer of the social knowledge, for it to become a trustable, accessible, useful public good \cite{N4}.

According to the mentioned World Economic Forum report \cite{Pentland2011}, personal data is becoming a new economic “asset class”, a key resource for the 21st century that will touch all aspects of society. The PNS vision aims at the creation of the scientific and technological framework that can make this new asset, the big social data, flow and become knowledge for all and everyone.

\section{Social sensing: state-of-art and research roadmap}
\label{socialSensing}

There are seven billion people on our planet today, together with five billion mobile phones and two billion PC's and laptops. For the first time in history, the majority of humanity is linked and has a voice. In prospect, though, the most striking change will come from the fact that these same mobile phones and other devices are increasingly sophisticated location-aware sensor platforms and their wireless networks support an increasing variety of sensors in cars, buses, and homes, besides providing Internet connectivity and full access to the Web and the social media. As a consequence, the mobile wireless infrastructure, in tandem with Internet and the Web, generates an ever richer and larger mass of signals and observations from reality that can be mined --- or ``reality mined'' --- in order to understand the patterns of human behaviour, monitor our environments, and plan the development of our society \cite{TechnologyReview2008}. This functionality is in its infancy at this point, but already these signals and abservations are being used to measure population flows into cities and slums, to map the movement of populations during emergencies, to identify neighbourhoods where social services are inadequate, to predict the outbreak of diseases, and to manage automobile traffic congestion \cite{Pentland2009}. We can expect that the evolution of this nervous system will continue at a quickening pace because of the continuous progress in computing and communication technologies. Networks will become faster, devices will become cheaper and will have more sensors, the resulting techno-social ecosystem will become ever more integrated, and the data science needed to model and understand human behaviour will become more accurate and detailed \cite{LP09}. In this paper, we use the term {\em social sensing} to denote the novel ICT paradigms needed to feed the planetary nervous system, i.e., the techniques for collecting the digital footprints generated by humans when interacting with the surrounding techno-social ecosystem and making such big data available for mining/analysis at properly high abstraction levels.

The complexity of social sensing derives both by the variety of forms of data left behind by the ICT mediated human activities/interactions, and by the variety of acquisition devices and methods available. Data formats range from text, graph, logs, streams, structured, and semi-structured data. Acquisition sources range from social networks, web logs, web content, links, transactions, search engines, social media etc. to the wide variety of smart devices spread in the environment (such as RFID readers, sensors and actuators, sensor-rich smart phones, or wireless communication technologies such as WiFi, Bluetooth and cellular networks).

In all cases, the low-level representation of human activities in sensed data makes it difficult to observe human behaviour. The major demand for social sensing is to complement the raw data with additional information conveying concepts of human behaviour and social semantics:
\begin{itemize}
\item either by automated annotation and enrichment methods that extract social semantics  along several dimensions from otherwise raw data,
\item or by advanced participatory systems, that explicitly involve people into the picture, by means of crowd computing, serious games, participatory sensing.
\end{itemize}

Such big picture may be mapped on four complementary research challenges: i) a toolkit of methods for semantic annotation of raw big data, ii) novel machine reading methods for learning knowledge bases from texts iii) analytical crawler to extract meaningful concepts from the social web, and iv) participatory sensing systems to engage people in effective and large-scale campaigns for collecting high quality data.

\subsection{Semantic Enrichment of Digital Breadcrumbs}
Digital breadcrumbs are essentially structured data, that record basic events underlying complex ICT applications: billing systems of TELCOs as well as resource allocation of the GSM network are based on simple CDR (Call Data record) registering position and duration of a phone call, accounting systems of credit cards or retails are based on simple cash transactions registering the cost of the bought items, car or personal navigation systems are based on GPS data recording the sequence of spatio-temporal position (instant, point),  web services are based on simple logs registering ip address, time and requests of web users, web search engines continuous improvement as well as personalization services  are based on simple query logs registering user id, time and keywords requested. Plenty of other examples might be done for sensors network and RFID based software systems.
These simple form of data, collected for many different purposes, contain historical and collective information of very basic human actions and they are stored for short or long time in efficient and enormous repositories or in smart personal devices, or they may fast pass trough a data streaming window.
To make such breadcrumbs usable for analysis it is needed to inject back the contextual knowledge of human action that got lost during sensing.  So discretizing, normalizing, aligning, sampling are preliminary steps towards making such data closer to the phenomena to be observed.   More properly, those are steps of a process aimed at adding context information to raw data by augmenting, tagging, and/or classifying data in relationship to dictionaries and/or other reference sources.

At a very basic level, enrichment conveys the idea that existing data is complemented with additional information and annotations that conceptualize salient aspects of data. Although we can expect that data will become richer and richer with explicit semantics tags, due to context-sensitive devices and possibly the semantic web, we have access today with big raw data, which need (and actually can) be processed in many ways so that the implicit semantics of users' activities can be exposed to some extent.

The literature is rich in proposals for semantic enrichment approaches and techniques, each one built on the background of some popular application domains and focusing on the acquisition of knowledge specific to the domain.

In case of mobility trajectories, i.e., sequences of time-stamped location records, the basic enrichment is the extraction of \emph{trips} obtained by segmenting each trajectory according to intervals of time when the object is considered as stationary (stop) and intervals when the object is moving (move). The characterization of stop/move episodes is domain dependent: stops and moves can be computed by taking into account several spatio-temporal criteria like position density, velocity, and direction.
Trips may be further conceptualized and annotated with contextual data extracted from application sources and from geographical sources as, for example, recording the goal of a person's trip to Paris (e.g., business, tourism), or inferring and/or recording the various transportation means used during a trip \cite{PRAABDGMPTY12}.
Another dimension is enriching trips with concepts related to the \emph{activities} performed during a day, representing the purpose of the movement. This implies the recognition of how stop/move patterns correlate to specific activities (home, work, shopping, leisure, bring/get,..). At the edge of current research \cite{Janssens2012,MYW12} the correlation is first learned from surveyed activity diaries and then used to classify trips of new users.

In case of query logs, the basic enrichment is the extraction of ``query sessions'', i.e. specific sets/sequences of queries submitted by a user while interacting with a search engine \cite{HG00}. Sessions represent the basic unit of information for tasks like query suggestion, learning to rank, enhancing user experience of search engine users, etc.  More advanced enrichment is to identify (conceptualize and label) \emph{tasks and missions} behind the user sessions extracted by the logs. This may be achieved by augmenting text in the query using space of concepts derived from Wikipedia/Wiktionary \cite{LOPST11}.

In case of tweets, enriching user profiles with concepts related to topics and entities implies to exploit prior social knowledge about bags of entities and categories, such as those provided by Wikipedia and DBpedia. At the edge of research, this mapping allows to crosslink the enhanced tweets to other sources such as news in \cite{DGL12} thus overcoming the vocabulary mismatch problem by mapping both tweets and news in the same entity-based coordinate system. The mixing of languages -- traditional idioms and tech idioms such as emoticons -- is another big challenge.

An advanced vision of semantic enrichment is that of a system that continuously and automatically enriches its own understanding of the context and content of the data it is receiving, by comparing it to its existing knowledge base and then building increasingly smarter classifiers upon that store.

More generally, the overall repertoire of semantic enrichment needs to be adaptive with respect to the continuous progress of sensing devices, to improve its capability to map human behaviour along multiple dimensions, and to become increasingly intelligent with use, by incorporating machine learning techniques.

\subsection{Human-level understanding of text at Web scale}

We need systems that progressively acquire the ability to interpret textual communications and to extract structured information (semantic objects) from various online sources, including text exchanged by people through the web, in pages, blogs, news, tweets, etc. This includes several sub-challenges including linguistic analysis in multiple languages, continuous learning and domain adaptation, dealing with temporal and spatial relations, etc..
The relatively new research field, denoted as Machine Reading \cite{EBC06} is the automatic, open domain, unsupervised understanding of text aimed at extracting an unbound number of relations on Web scale. Machine Reading is aimed at providing the human-level understanding of documents that is instrumental to perform social analysis of knowledge that combines information present in individual documents with relations among documents and people using those documents. MR is based on combined natural language processing (NLP) and statistical techniques. Continuous learning is fundamental to deal with open-domain acquisition from disparate sources, for collecting and contrasting evidence at various level of linguistic knowledge: terminological and syntactical. Discriminative analysis is needed to recognize emerging syntactic patterns.

MR goes beyond document retrieval systems and search towards systems capable of understanding and predicting user behaviour. Extraction of knowledge from texts has been attempted initially with simple text mining techniques, which exploit recurrences in the data and extract rules to represent those recurrences. Later machine learning techniques have been deployed. These however typically require substantial amount of human effort in manual engineering and/or labelling examples. As a result, they often focus on restricted domains and limited types of extractions (e.g., a pre-specified relation). TextRunner \cite{BCSBE07} is a notable attempt at ``open domain information extraction'' \cite{BE08} that shifts from the traditional supervised paradigm, whose costs are prohibitive at web-scale.

The Never-Ending-Language-Learning (NELL) system \cite{MBCHW09} starts from an ontology with a few example instances for each concept, and learns to extract more instances by reading new text. It repeatedly labels new examples and retrains the extractor, with a novel addition that enforces the mutual exclusive constraint among non-overlapping concepts and substantially improves extraction accuracy.

Though MR meets the challenge of scalability and domain independence, it still has problems, notably an often large number of incoherent and uninformative items extracted. A promising solution is described in \cite{PD10}, which aims at abstracting over lexicalized strings in an unsupervised manner. This methodology, however, fails to meet the web-scale target, because the number of required operations at each iteration grows rather quickly. An alternative approach is that of building taxonomies over the text, using a graph-based approach \cite{NVF11}.

Another interesting new problem that has emerged in this area is that of identifying contradicting opinions \cite{TP11}, which is an effective way for focusing on the interesting parts of the data. For example, we identify that suddenly the opinions of people on some topic, such as nuclear power, have changed from positive to negative, or that a new popular topic, such as fiscal reforms, is causing both positive and negative sentiments. These results could be the basis for the further analysis of the data, leading to a better understanding of the causes, as well as the future social trends.

\subsection{Analytical crawling}

Analytical crawling improves current technology by adopting a goal driven approach capable to combine semantic enrichment and machine reading techniques to the aim of  gathering semantic-rich data linked across diverse web resources.

Crawling techniques so far has been designed to support search tasks and the research mainly focus on improving indexing structure both in efficiency and efficacy, the new challenge will be to push part of the intelligence of social sensing as far as possible into the gathering process and make it capable of constructing, maintaining and using multi-dimensional network of semantic objects.

A simple example is the study of relationships of very important persons all over the World: this might be the network constructed by crawling their co-occurrence on the same article on news, blogs etc. bound by positive or negative terms, phrases, or opinions. Another example is the study of relationships among international institutions and organizations: again this might be the network constructed by crawling their co-occurrence on the same article on news, blogs, etc., bound by collaboration ties. The crawling process will construct new nodes or new edges annotated with relevant concepts by analysing on the fly the web content at hand, extracting and conceptualizing the concepts by matching with the current network, with external sources and with the analytical goals. The crawling process will continue by following pertinent URLs, selected on the basis of the semantic network constructed incrementally; such network, at the end of the process, is the final output and input for the subsequent social mining analyses.

\subsection{Participatory sensing}

Another complementary big challenge for social sensing is to develop methods that are able to directly involve people as active/passive actors in the data collection phase.
The objective is to develop tools to elicit facts, opinions and judgment from crowds.
Participatory sensing aims at building, on demand, a sensor network infrastructure (as for example the mobile phone network, or wearable electronic device network or a social network) based on people collaboration towards a common sensing task (crowdsourcing).  There might be different levels of user involvement in participatory social sensing, ranging from a mere explicit donation of personal digital traces in sophisticated survey campaigns, to actively perform subtasks useful to the overall sensing. The sequel gives a briefly overview of such different forms of user involvement \cite{N4}.

\paragraph{Ad hoc campaigns.}Survey campaigns, the basic instrument of data collection for data analysts, statisticians, social scientists etc., has been completely renovated by the proliferation of ICT technology: from web, to smart phones to sensor networks. We can imagine setting up ad hoc survey campaigns to sense nearly any kind of human behaviour. Smart phones are regularly used for surveys in transportation engineering and mobility analysis to engage group of volunteers in keeping track or their travel diaries. The respondents essentially annotate their positions with the kind of activity performed, typically w.r.t some predefined activity ontology, which is incorporated in the mobile application \cite{Janssens2012,JCE12}. The incentive models for scaling this kind of campaigns up to large populations is a hot research issue among travel demand researchers.

As far as social interactions are concerned, while social web applications generate copious amounts of data on how people behave and interact with each other online, many characteristics are expressed only in real-world, face-to-face interactions. In order to capture these behaviours, innovative methods and tools have been proposed, that can accurately capture face-to-face interactions. One example is the \emph{sociometric badge} (commonly known as a ``sociometer''), a wearable electronic device capable of automatically measuring the amount of face-to-face interaction, conversational time, physical proximity to other people, and physical activity levels using social signals derived from vocal features, body motion, and relative location. At MIT Media Lab \cite{Pentland08}, several hundred sociometric badges have been built and experimented in campaigns at real organizations to measure individual and collective patterns of behaviour, to predict human behaviour from unconscious social signals, identify social affinity among individuals working in the same team, and enhance social interactions by providing feedback to the users.

\paragraph{Human Computation and Crowdsourcing.}The social web has prompted the emergence of a new computation paradigm, called Human Computation, applied in business, entertainment and science, where the interaction among users is harnessed to help in the cooperative solution of tasks. According to \cite{QB11}, a system belongs to the area of Human Computation when human collaboration is facilitated by the computer system and not by the initiative of the participants. A classical example of HC is content processing for multimedia search applications. In this domain, the goal is automatically classifying non-textual assets, audio, images, video, to enable information retrieval and similarity search, for example, finding songs similar to a tune whistled by the user or images with content resembling a given picture. Recognizing the meaning of aural and visual content is one of the skills where humans outperform machines and it is now commonly recognized that multimedia content analysis can benefit from large scale classification performed by humans; applications like Google Labeler submit images from a large collection to human users for receiving feedback about their content and position, which can be integrated with machine-based feature extraction algorithms.

Crowdsourcing is a facilitator of human computation: it addresses the distributed assignment of work to a vast community of executors in a structured platform \cite{Howe06}. A typical crowdsourcing application has a Web interface that can be used by two kinds of people: work providers can enter in the system the specification of a piece of work they need (e.g., collecting addresses of businesses, classifying products by category, geo-referencing location names, etc); work performers can enrol, declare their skills, and take up and perform a piece of work. The application manages the work life cycle: performer assignment, time and price negotiation, result submission and verification, and payment. Examples of crowdsourcing solutions are Amazon Mechanical Turk and Microtask.com. Application areas are the most varied: speech transcription, translation, form filling, content tagging, user evaluation studies are a few examples. For example, it is possible to develop mechamisms for users with common interests in a certain task being done, that allow to distribute work in a lie-tolerant way.

\paragraph{Games with a purpose.} Human Computation can assume a variety of forms, according to the scale at which humans are engaged, the tasks they are called to solve, and the incentive mechanisms that are designed to foster participation \cite{QB11}. Games with a Purpose (GWAPs) is a line of work that focuses on exploiting the billions of hours that people spend online playing with computer games to solve complex problems that involve human intelligence \cite{Ahn06,LA09}. The emphasis is on embedding a problem-solving task into an enjoyable user experience, which can be conducted by individual users or by groups; several game design paradigms have been studied \cite{LA09} and the mechanics of user's involvement has started being modeled formally.

At the edge of research we have today the idea of techno-social systems for Community and Campaign Management, which go far beyond the current crowdsourcing systems such as CrowdDB \cite{Franklin2011}, Turk \cite{Marcus2011}, and Snoop \cite{PP2011}. These systems are based oh hiring vast populations of un-ware executors to perform simple tasks for small economic incentives. On the contrary, analytical crowdsourcing systems should involve vast populations of aware participants, which perform classical social responses, such as liking, ranking, and tagging, organized into social networks and collaborating communities, within incentives models that foster the emergence of collaborative behaviour \cite{HY2009}.  An interesting example is the DARPA’s Network Challenge, where MIT researchers \cite{TCGKKW11} harness online crowds to beat Darpa Balloon Challenge in Just 9 Hours.  It is a bizarre national balloon hunt that experts initially said was impossible but was accomplished in nine hours. In that challenge, MIT researchers devised a payment system that rewarded accurate balloon-finding tips as well as the people who invited the finders. That contest was worth a more handsome \$40,000, however. In April 2012, Manuel Cebrian, and his team won the Tag Challenge, a new State Department-sponsored competition to find five fake jewel thieves in five countries within $12$ hours.
This enhanced view of participatory sensing is clearly interconnected with the idea of a Global Participatory Platform, fostering global awareness and collective intelligence, discussed in \cite{Shum12}. The vision is that aware participation in a collective sensing campaign will enable the creation of social data of higher quality and realism, capable of portraying concurrently different dimensions of social life. Finally, the ability of combining data sensed both by opportunistic and participatory methods is expected to open new avenues for learning systems, capable of recognising and mapping human behaviour at global and local scale.

\section{Social Mining: state-of-art and research roadmap}

Social Mining aims to provide the analytical methods and associated algorithms needed to understand human behaviour by means of automated discovery of the underlying patterns, rules and profiles from the massive datasets of human activity records produced by social sensing. The patterns and models created by the Social Mining methods and algorithms will provide the backbone of the Planetary Nervous System, i.e., the building bricks for the analyses and simulations of the Living Earth Simulator \cite{N2}. Although data mining and statistical learning from traditional databases are relatively mature technologies \cite{KumarBook,StatisticalLearning}, the emergence of big data of human activities, their networked format, their heterogeneity and semantic richness, their magnitude and their dynamicity pose new exciting scientific challenges.

We need significant advances in providing large-scale data mining of digital and physical traces generated across social media, environmental, mobile and wearable sensing devices in order to predict human behaviour and diffusion processes in highly complex and heterogeneous socio-technical systems in different domains (e.g., collective action, market and financial issues, healthcare, diffusion of  technology and innovation, cultural and educational issues).

Human Data Mining, Reality Mining: these are the new keywords witnessing the flourishing of novel data mining, statistical learning and network science methods centred on the digital footprints of human activities. The following fields are the ground where the different disciplines are converging.

\paragraph{Social network analysis} refers to the study of interpersonal relationships, with the purpose of understanding the structure and the dynamics of the fabric of human society. It is currently a very hot and attractive research area with increasing presence in the major journals and conferences of the aforementioned disciplines. Roughly, we may distinguish two major lines: the statistical laws regulating statics and dynamics of complex networks \cite{WS98,BA99,Caldarelli2007,Newman10,EK10}, and the methods aimed at discovering patterns, evolutionary rules, community structure and the dynamics in large social network: i) community discovery \cite{Fortunato2010,CGP11}; ii) information propagation \cite{K00,KKT03,PV01,KE05}; iii) link prediction \cite{LK03,KKYST09,LHK10}; iv) temporal evolution \cite{LKF05,HS11,MRMPO10,BCGMP10,BBBG10}; v) multilevel networks \cite{GBSH12,BCGMP11A,TL09}.

\paragraph{Social media mining} methods for mining from heterogeneous data, especially text,  sensed from different on line sources (tweets, mails, blogs, web pages, link structures etc.) to the purpose of extracting the hidden semantics from them. A key topic is \emph{opinion/sentiment mining}, \cite{PL08} which has come to play a key role in customer relationship management, consumer attitude detection, brand/product monitoring, and market research in general \cite{ES10}. Interest in these applications has spawned a new generation of companies and products devoted to online reputation management, market perception, and online content monitoring in general. Historically, one of the most important incarnations of opinion mining has been sentiment classification, the task of classifying a given piece of natural language text (be it a short remark, or a blog post, or a full-blown product review) not according to the topic it is about (as in standard text classification) but according to the opinions expressed in it.

\paragraph{Mobility data analysis} emerged as a vital field, leveraging the spatio-temporal dimensions of big data (such as mobile phone call records and GPS tracks, generated by current mobile communication technologies and wireless networks) to the purpose of understanding human mobility behaviour, evolutionary patterns, daily activity patterns, geographic patterns. Network scientists concentrated on the models of human movements, showing the emergence of power laws both in the probability of a banknote or a person to traverse a certain distance, and in the distribution of people's radius of gyration, the characteristic distance travelled by a person \cite{BHG06,GHB08}. Heavy-tailed distributions also arise from the number of distinct location visited by humans and from the visitation frequency, that is the probability of a user to visit a given location \cite{SKWB10}.  Models of pedestrian behavior have also been developed and validated on empirical data, capable to predict crowd disasters \cite{MHT11}.
Computer scientists put forward {\em mobility data mining}, aimed at discovering patterns of interesting mobility behaviors of groups of moving objects \cite{GP08}. Examples of this process are the mining of frequent spatio-temporal patterns, trajectory clustering, and prediction of locations and trajectories \cite{TPNG11,GNPP07,GNP+11,MPTG09}.  A recent trend in research focuses on extracting mobility profiles of individuals from their traces, trying to address one of the fundamental and traditional question in the social science: “how human allocate time to different activities as part of the spatial temporal socio-economic system”. By profiling individuals according to their daily activities, the ultimate goal is to provide a picture of how group on individuals interact with different places, at different time of the day in the city \cite{TPNG11,JFG12,FM12,AZSR02}.

\paragraph{}A definite research trend, although in its infancy, is the convergence of data mining and social network analysis towards the definition of novel mining models and patterns that combine the different dimensions of human activities: social ties, mobility, opinion and preferences, socio-economic behaviour etc.: What is the interplay between the social and mobile network? How can we automatically identify strong and weak ties in virtual human relationships by analysing the human interaction patterns? How can we use the social and mobile networks to answer question about the socio-economic structure of a city? We expect to observe an explosion of social mining methods along this path, as richer and more comprehensive data will be harvested by social sensing. An example of early work in this direction is \cite{WPSGB11}.

With the size and complexity of big data, also new computational challenges will emerge. People have been working with graphs for hundreds of years, but the graphs now being plotted from social networks or sensor networks are of an unprecedented scale, they are produced continuously and massively distributed.  Dealing with graphs of that size and scope, and applying modern analytic tools to them, calls for better models, scalable algorithms and other innovations.

We distinguish two main complementary research directions for social mining, focussing on (i) the invention of methods for the extraction of high-level models and patterns of human behaviour, and (ii) the engineering of such methods in the complex scenarios of techno-social systems at global scale.

\subsection{Mining Patterns of Human Behaviour}
Social mining is the problem of discovering patterns and models of human behaviour across the various social dimensions: it is aimed at extracting multi-dimensional patterns and models from multi-dimensional data. To this aim, social mining needs novel concepts and techniques of \emph{multi-dimensional pattern discovery} and of \emph{multi-dimensional social network analysis}, whose ultimate goal is to understand human behavior:
\begin{itemize}
\item the discovery of individual social profiles,
\item the discovery of collective behaviors,
\item the analysis of spreading and epidemics over multi-dimensional networks,
\item the analysis of sentiment and opinion evolution.
\end{itemize}
The key point of social mining is to combine the macro and micro laws of human interactions within a uniform multi-dimensional analytical framework, encompassing both the global models developed by complex network theory and the local patterns discovered by data mining in the sensed data.
Novel social mining techniques that are needed include the discovery of groups/communities that share similar socio-economic behaviour, of behavioural patterns of temporal evolution, of individuals and communities that provide weak early signals of change, of correlation patterns/behaviours across different domains.

\paragraph{Multi-dimensional Social Network Analysis.}Social sensing is providing data in form of networks with rich semantics, content and metadata attached to nodes and edges. We need the foundations of a multi-dimensional network analytics, capable of extending the methods and tools of network science to the multi-layered, interconnected networks resulting from the novel social sensing techniques. The needed concepts and tools will exploit not only the network topology but also the available social metadata. Examples include: (i) multi-dimensional, evolutionary community detection and link prediction driven by similar socio-economic behavior (homophily), (ii) definition of multi-dimensional measures of network centrality, relevance and trust; (iii) correlations analysis between diverse social dimensions, e.g., proximity metrics in the evolving network of physical contacts derived from user mobility traces and proximity metrics defined with respect to the social space of people profiles.
\paragraph{Multi-dimensional Social Pattern Discovery.}A central method for data mining is pattern discovery: the automated extraction of regularities, or analogously of segments of the whole populations whose members exhibit a common property. We need to expand pattern discovery towards the emergent patterns and rules describing socio-economic behaviors of significant sub-populations. Examples include: (i) discovery of correlation patterns connecting behaviours across different social dimensions; (ii) multi-dimensional evolutionary pattern mining from social and mobility networks, from streaming data with uncertainty; (iii) discovery of evolutionary patterns for trust and risk measurement.

\begin{figure}\centering
\resizebox{0.75\columnwidth}{!}{%
\includegraphics{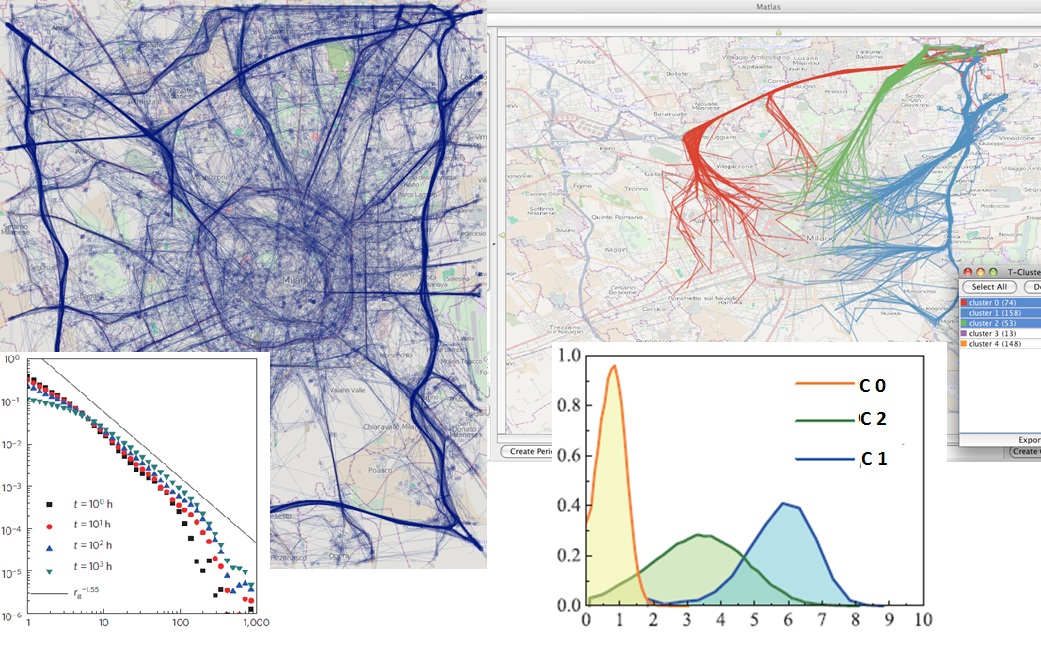} }
\caption{The GPS trajectories of tens of thousand cars observed for one week in the city of Milan, Italy, and the power-law distribution of users’ radius of gyration and travel length (left); the work-home commuting patterns mined from the previous dataset by trajectory clustering and the normal distribution of travel length within each discovered pattern (right).}
\label{trajectories}       
\end{figure}

\paragraph{Combining network science and pattern mining/inference.}Social mining pushes towards the convergence of the complementary strengths and weaknesses of network science and data mining \cite{N12,N13}. The former is aimed at discovering the global models of complex social phenomena, by means of statistical macro-laws governing basic quantities, which show the behavioral diversity in society at large. Data mining is aimed at discovering  local patterns of complex social phenomena, by means of micro-laws governing behavioral similarity or regularities in sub-populations. The dualistic approach is illustrated in Figure \ref{trajectories}. In the overall set of individual movements across a large city we observe a huge diversity: while most travels are short, a small but significant fragment of travels are extraordinarily long; therefore, we observe a long-tailed, scale-free distribution of quantities such as the travel length and the users’ radius of gyration \cite{GHB08}. Mobility data mining can automatically discover travel patterns corresponding to set of travellers with similar mobility: in such sub-populations the global diversity vanishes and similar behaviour emerges \cite{GNP+11}. The above dual scenario of global diversity (whose manifestation is the emergence of scale-free distributions) and local regularity (within clusters, or behavioural profiles) is perceived today as the signature of social phenomena, and seems to represent a foundational tenet of computational social sciences \cite{N14,LP09}. Although network science and data mining emerged from different scientific communities using largely different tools, we need to reconcile the macro vision of the first with the micro vision of the second within a unifying theoretical framework, because each can benefit from the other and together have the potential to support realistic and accurate models for prediction and simulation of social phenomena.

\paragraph{Analytical challenges for social mining.}Based on the novel social mining techniques, it will become possible to address in a systematic way many challenging questions about multi-dimensional social phenomena, ranging from the individual sphere of personal behavioral profiles, to the collective sphere of group or subpopulation behavior, to the global phenomena related to diffusion in the social network, to the analysis of how sentiments and opinions vary in our societies.
\begin{itemize}
\item \emph{Individual profiles}. Mining the multi-faceted personal data of each individual citizen allows to discover personal routines, or profiles. Considering the entire collection of profiles over a population yields an unprecedented opportunity to map the social behavior at a realistic scale. While profiling has been up to now a key ingredient of commercial systems for target marketing and recommendation, its generalization for social mining has the potential of boosting the comprehension of social behaviour, in that it represents a trade-off between the oversimplification of global measurements and the excessive diversity of individual behaviors.  Exemplar lines of investigation on profiles include: (i) mining of personal mobility-activity diaries to create realistic profiles and routines, allowing to characterize systematic vs. non-systematic activity and mobility; (ii) Multi-dimensional profiling, combining mobility, social ties, interests, life-style and shopping behaviour, based on the full variety of data harvested by social sensing; (iii)  profiling of social network users w.r.t. epidemic spreading and diffusive processes across different media.
\item \emph{Collective patterns and models}. Multi-dimensional mining methods applied at a global scale over entire populations (and/or their individual profiles) allow to discover and analyze patterns and models pertaining to interesting sub-populations, where the emergence of collective behavioural regularity highlights the local laws governing aspects of social life in different communities. Exemplar lines of investigation on profiles include: (i) models of the interplay among the social dimensions in different communities; (ii) dynamic patterns and predictive models of social evolution; (iii) discovery of the geographical and temporal extension/evolution of collective patterns; (iv) discovery of shopping behaviour and life-style patterns from large-scale market transaction data; (v) discovery of the patterns of multi-dimensional network evolution.
\item \emph{Multi-dimensional diffusion processes}. The models of epidemics and diffusion processes need to be generalized to multi-dimensional social networks, in order to define realistic models of social contagion, information diffusion and cascading behavior. Exemplar lines of investigation on profiles include: (i) patterns of virality and its impact on crowd computing; (ii) analysis of information diffusion in evolving networks of social contacts, impact of multi-dimensionality on diffusion speed; (iii) analysis of virality (of messages, contents) and leadership (of users or groups in a network).
\item \emph{Aggregated sentiment analysis and opinion mining}. Learning and monitoring the evolution of collective sentiments and opinions from user generated content in social media. Exemplar lines of investigation on profiles include: (i) deep analytics of sentiment-based contradictions, discovery of patterns and indicators of sentiment evolution and conflicting opinions, handling of concept drift by means of incremental learning; (ii) analysis of emotions and mood emerging from dialogues; (iii) diffusion phenomena across different media: when a trendy topic/hashtag in Twitter can likely become an argument well covered by online news in the next future?
\end{itemize}

\subsection{Computational aspects of social mining}
Although data mining from traditional databases is a mature technology, the big data from human activity, their networked nature, heterogeneity and semantic richness, combined with magnitude and dynamicity, pose new exciting scientific challenges. The mainstream paradigm for analysis today is: gather data first, and analyse them later. This paradigm falls short in view of the challenges posed by the PNS, which requires for real-time reactions to changes in the techno-social ecosystem. This is a perspective that calls for fundamental advances in many fields of data mining and machine learning, such as: concept drift and management of model evolution/obsolescence, learning from continuous data streams, real-time and interactive analytics and learning (putting users within the loop), massively distributed data mining, high-performance data mining and big graph analytics.

\begin{itemize}
\item \emph{Concept drift \& streaming data mining}: we need a general framework for the on-line monitoring and analysis of high-rate evolving data streams. The goal of this framework is to provide a declarative process for the development of complex analytics applications over streaming data, by handling uncertain and inexact data and providing mechanisms for correlating and integrating real-time data (fast streams) with historical records \cite{MZK05}. A key issue here is how to deal with concept-drifting, i.e., how modelling can self-adjust in real time as the characteristics of the continuously streaming data change in time and the mined patterns and models become obsolete. Here, the large body of research in active, online and reinforcement learning plays a central role, albeit it needs to scale up to big data complexity and size.

\item \emph{On-line ingestion of semi-structured and unstructured data}: Techniques for the effective analysis of streaming non-structured data (e.g., streaming news feeds, continuous updates on tweets, blogs, product reviews, etc.). These techniques aim at tapping into the wealth of information outside the realm of structured data, identifying emerging trends, discovering similar and diverging cultural signals, uncovering hidden social cliques, and sensing the pulse of different societal groups, all in real time.

\item \emph{Massively distributed data mining}: Data storage and data mining protocols that minimize the amount of data moved over the network, favouring the delocalization of the computational efforts toward the data rather than vice versa;

\item \emph{High performance data mining and big graph analytics}: methods to effectively store, move and process huge amounts of data: in particular new algorithmic paradigms (such as map-reduce) should be defined and specific mining techniques should be introduced and implemented according to these paradigms, as well as data engineering methods for large networks with up to billions nodes/edges including effective compression, search and pattern matching methods.

\end{itemize}

\section{Trust Networks and Privacy-aware Social Mining: state-of-art and research roadmap}
The perspective of a Planetary Nervous Systems based on social sensing and mining sheds a radically new light on the questions of personal data, privacy, trust and data ownership. Before the “big data revolution”, the protection of the private sphere of individual persons against the intrusion threats made possible by the ICTs has been the main focus of regulations and public debate. Now that the importance of personal data as a source of societal knowledge is becoming evident and urgent, the matter reconfigures more maturely as a trade-off between conflicting, legitimate interests: on one side, the right of people to keep control on their private space and, on the other side, the right of people to access collective knowledge as a public good \cite{N21}.

One of the toughest challenges posed by the new ability to sense and analyze the pulse of humanity is creating what has been called a {\em new deal on personal data}, which implies a change of perspective on the issues of privacy, trust and data ownership \cite{Pentland2011}. Advances in analysis of human activity data must be approached in tandem with a better understanding of how to create value for the producers and owners of the data within a framework that favors the development of trust and the control over own personal data, while at the same time protecting the public good. We must avoid both the retreat into secrecy, so that big data become the exclusive property of private companies and remain inaccessible to the common good, and the development of a “big brother” model, with government using the data but denying the public the ability to investigate or critique its conclusions. Neither scenario will serve the long-term public interest in a transparent, accessible and efficient knowledge society.

The use of anonymous data should also be enforced, and analysis/mining at collective level should be preferred over that at individual level. Solid models of data sharing need to be developed, capable of taking into account both the privacy expectations of consumers as well as businesses’ legitimate interests. Most personal data are being collected by private organizations – location records, financial transactions, public transportation, phone and Internet communications, query and navigation logs, and so on. Therefore, companies have a crucial role, and so either extensive government regulation or market mechanisms with auditing will be needed in order to incentivize/push owners to set free the power of the data they hold. However, the crucial problem is how to make each individual user fully aware of the value of her own continuously generated personal data, and how to empower each user’s ability to extract knowledge from her own data, as well as to exercise control, usage, sharing, disposal over her own data.

The choice is to design awareness and incentive mechanisms that have sufficient accountability to guarantee individual and societal safety, but how to achieve this goal is still largely an open question, both at regulatory and technological level. It is apparent, though, how this debate is becoming mainstream, as also witnessed by the reform of the European data protection directive, recently proposed by the European Commission\footnote{European Commission proposes a comprehensive reform of the data protection rules, 25 January 2012, see \url{http://ec.europa.eu/justice/newsroom/data-protection/news/120125_en.htm}}, which introduces concepts such as data portability among different service providers, data oblivion, user profiling, and privacy-by-design. We now briefly discuss the research challenges behind (i) trusted networks for social sensing that foster users to share personal data within a transparent framework, and (ii) privacy-preserving methods for social mining that enable the analysis of personal data in a safe framework.

\subsection{Trust-aware social sensing and the new deal on data}
The first step toward creating a fair information market is to give people ownership of the data that is about them \cite{Pentland2012}. Just as with financial and commodity markets, the first step toward a healthy market is creation of an asset class such as land rights. This is why the World Economic Forum publication is subtitled `Emergence of a New Asset Class' \cite{Pentland2011}.

Quoting Pentland \cite{Pentland2012}, the simplest approach to defining what it means to {\em own your own data} is to go back to Old English Common Law for the three basic tenets of ownership, which are the rights of possession, use, and disposal:
\begin{enumerate}
\item You have a \emph{right to possess your data}. Companies should adopt the role of a Swiss bank account for your data. You open an account (anonymously, if possible), and you can remove your data whenever you’d like.
\item You, the data owner (or data subject in the language of the European directive on data protection), must have \emph{full control over the use of your data}. If you’re not happy with the way a company uses your data, you can remove it. Everything must be opt-in, and not only clearly explained in plain language, but with regular reminders that you have the option to opt out.
\item You have a right to \emph{dispose or distribute your data}. If you want to destroy it or remove it and redeploy it elsewhere, it is your call.
\end{enumerate}

Individual ownership needs to be balanced by the legitimate interests for corporations and governments to use personal data – credit card numbers, home addresses, etc. – to run their  operations. The “new deal on data” therefore gives individuals the right to own and exercise control on a copy of the data about them rather than ownership of corporations’ internal data, as long as that data is required for legitimate operations, transparently known by the data owner (or data subject, i.e., the individual that is being portrayed in the data). On top of this, the data owner has the power to withdraw her personal data from the data provider at any time, or have it forgotten after a reasonable amount of time. The newly proposed reform of the EU data protection directive follows this path of reasoning.

Creating a trust ecosystem where this new deal can take place is another big challenge. Enforcement is not simply authenticating the identity of an individual, but also validating whole series of “claims” and “privileges” an individual, institution, or device may make, that give them access to highly-valued services and knowledge resources. As more and more business, financial, civic and governmental services use personal data, the integrity and interoperability of a global authentication and “claims” infrastructure will be paramount. Clearly, such global infrastructure, like the Web itself, cannot be a centralized authority, but rather a highly distributed and user-centric, self-organizing ecosystem. Essentially, we should mimick at the digital level the social mechanisms that, in real life, better promote rapid innovation and self-correction and favor ethical/fair behavior while keeping crime at a physiological, sustainable level. Similar to the current Open Identity Exchnge (OIX) ecosystem, the advocated `Trust Network’ will need to continuously monitor, flag and pursue fraudulent and deceptive behaviors. This  requires not only innovations in ICT systems, but also in policy and contract law. The Law Lab at the Harvard Berkman Center has developed a prototype of such a trust network, in collaboration with the MIT Human Dynamics Lab \cite{Pentland2009}. As it is reasonable to assume that verifiable trust reduces transaction risks and builds customer loyalty, it will be in the economic interest of the companies offering future online and mobile services to become verifiable trusted stewards of personal data.

\subsection{Privacy-preserving social mining and privacy-by-design}
Anonymity as a form of data protection is increasingly difficult to achieve, as ever-more detailed data are being collected. This results in the fact that anonymization simply cannot be accomplished by de-identification (i.e., by removing the direct identifiers contained in the data). Many examples of re-identification from supposedly anonymous data have been reported in the scientific literature and in the media, from health records \cite{SS98,Sweeney02,NYT06} to query-logs  to GPS-trajectories. Despite this negative context, data protection technologies have come a long way: since the seventies there has been active research in statistical disclosure control and since the late nineties the related discipline of privacy-preserving data publishing and mining has also taken off. Many subtle threats to privacy and data protection rights have been found, and several methods for creating anonymous or obfuscated versions of personal datasets have been proposed, which essentially look for an acceptable trade-off between data privacy on one hand and analytical utility on the other hand.
The first privacy dimension is related to the personal sphere, namely individual privacy. In fact, models and patterns are developed on the basis of human activity data such as mobile phone call records, which contain private information concerning the places visited by each individual user and personal relationships with other users, and may reveal sensitive personal traits. The second dimension of privacy is related to the business and institutional sphere, namely corporate privacy. In the case that the modeling is performed cooperatively among several companies or institutions, or outsourced to third parties or cloud services, each individual carrier would like to protect not only the personal data of their customers, but also the corporate strategic knowledge represented by the data, e.g. the profiles of their customers.

In the literature, a variety of techniques have been put forward to tackle different cases \cite{AY08,BF10}:
\begin{itemize}
\item \emph{privacy-preserving data publishing} methods, aimed at the safe disclosure of transformed data under suitable formal safeguards against re-identification of individuals represented in the data (e.g., anonymization, randomization, etc.) \cite{Samarati01,MKGV06,GWY07,XT06};
\item  \emph{privacy-preserving knowledge publishing} methods, aimed at the safe disclosure of analytical models under suitable formal safeguards of individual or corporate privacy, as extracted patterns, rules, or clusters may reveal sensitive information in certain circumstances \cite{ABGP08} ;
\item  \emph{knowledge hiding} methods, aimed at the disclosure of data transformed in such a way that certain specified secret patterns, hidden in the original data, cannot be found \cite{VEBS04};
\item  \emph{secure multiparty analytics over distributed databases}, aimed at the analysis of datasets that are partitioned and distributed among several parties that do not want to (or cannot) share the data or certain corporate information that is represented in the data, but are interested in developing global models of common interest \cite{KC04,GSW05};
\item  \emph{privacy-preserving outsourcing}, aimed at enabling analytics performed by third-parties (e.g., cloud services) over corporate datasets, using encryption techniques to protect the input data and, if needed, the output models \cite{WCHKM2007,GLMPW2011};
\item  \emph{differential privacy}, a general-purpose privacy notion which enables analytics performed by third-parties (e.g., cloud services) over corporate datasets while guaranteeing  individual privacy \cite{DMNS06,Dwo06}. Informally, a randomized function of a database is differentially private if its output distribution is insensitive to the presence or absence of any particular record in the database. Therefore, if the analyses allowed on a database are guaranteed to preserve differential privacy, then the protection of individual records is preserved. Up to date, the problem of this notion is its tendency to compromise data quality.
\end{itemize}
A key approach for privacy-preserving data mining is privacy-by-design \cite{Monreale2011}, i.e., the idea of inscribing privacy protection into the analytical technology by design, so that the analysis incorporates the relevant privacy requirements from the very beginning, \cite{Monreale2010} offers a instance of privacy-by-design in the case of personal mobility trajectories, obtained from GPS devices or cell phones. The results show how such trajectories can be anonymized to a high level of protection against re-identification, while preserving the possibility of mining sophisticated patterns of human mobility, which enables novel powerful analytic services for info-mobility or location-based services.
In this research, the linking attack is considered, i.e., the ability to link the published data to external information, which enables some respondents associated with the data to be re-identified. In a trajectory linking attack, the malicious party M knows a sub-trajectory of a respondent R (e.g., a sequence of locations where R has been spied on by M) and M would like to identify in the data the whole trajectory belonging to R, i.e., learn all places visited by R. How can we guarantee that the probability of success of this attack is very low while preserving the utility of the data for meaningful analyses? Consider the personal trajectories represented in Figure \ref{trajectories}(left). Each trajectory is a de-identified sequence of time-stamped locations, visited by one of the tracked vehicles. Albeit de-identified, each trajectory is essentially unique --- very rarely two different trajectories coincide, given the extremely fine spatio-temporal resolution. As a consequence, the chances of success for the linking attack are not low; if the attacker M knows a sufficiently long sub-sequence S of locations visited by the respondent R, it is possible that only a few trajectories in the dataset match with S, possibly just one. In practice, guessing the home place and the work place of most respondents is rather easy. Assume that we want to discover the groups of trajectories that share common mobility behavior, such as the commuters that follow similar routes in their home-work trips (Figure \ref{trajectories}(right)). A data-driven anonymizing transformation of the trajectories can be defined, where trajectories are generalized from sequences of points to sequences of cells, so that the probability of re-identification drops significantly, obtaining a safe theoretical upper bound for the worst case (i.e., the maximal probability that the linking attack succeeds), and an extremely low average probability. In the example, the probability of success is bounded by 0.05, and it is below $10^{-3}$ for 95\% attacks. Moreover, the commuting clusters obtained from the anonymized data essentially coincide with those obtained with the original data, demonstrating that the desired quality of the analytical results can be achieved in a privacy-preserving setting with concrete formal safeguards, where the protection with respect to the linking attack can be measured. The impact of this method is far reaching: once an analytical process has been found and specified, it can be deployed and replicated with the mentioned privacy-preserving safeguards in order to find mobility clusters in different periods of time, in different cities, in different contexts: once deployed, it is a safe service that generates knowledge of the expected quality starting from measurably anonymous data.

\section{Our innovative approach}

In our FuturICT vision, the Planetary Nervous System is the techno-social ecosystem in charge of collecting, processing and interpreting big data from all available sources, built on top on privacy-aware social sensing and mining. Efficient methodologies and techniques through which the pulse of society and the awareness of relevant global phenomena can be detected and inferred from the data will be a key part of this work.

This social knowledge discovery process will be driven by high-level analytical goals provided by the Living Earth Simulator (e.g. “give an assessment of consumer confidence in Asia”), and based on an integrated ontology of large-scale socio-economic systems, accounting for interactions and value-exchange processes involving humans, organizations, machines, norms, and the physical environment, which will provide the background knowledge needed for massive social sensing and mining. Such ontology will be populated by the data analysed by the PNS, resulting in a distributed knowledge base capturing the “living Earth’s social state” that will be a basis for the social awareness (and from there the social adaptability) of the system.

Social knowledge and social awareness will also be exploited for developing novel self-* and social-inspired paradigms for designing a resilient and social-informed global nervous system. As discussed in our social sensing approach, data collection will be performed both in the cyber and in the physical/real world. Cyber-world data collection will exploit data mining techniques to extract relevant data from online sources. On the real-world side, participatory and opportunistic data collection will be encouraged: users’ personal devices will be like a probe in the real world collecting relevant data both directly (exploiting on-board sensors) and indirectly (by exchanging context- and social-aware data with their peers). Self*-policies will drive the devices behaviour to autonomously take appropriate actions. To preserve the users control on user-generated data and digital footprints, collected data will be pre-processed and aggregated at the user level, driven by the high-level goals and social-awareness, and exploiting opportunistic networking/computing techniques.

Privacy-by-design will be adopted as a basic approach, in such a way that the protection of personal data will be inscribed into the collection, analysis and mining processes of the global nervous system. Making data anonymous, i.e., achieving a negligible probability to re-identify the specific persons portrayed in the data sensed by the participatory platform, will be the systematically pursued with powerful privacy-preserving data publishing, analysis and mining technologies.

Based on such goals, the operation of the Planetary Nervous System requires the following basic functionalities:
\begin{enumerate}
\item Social sensing and social mining techniques that enable the efficient collection and interpretation of massive data from diverse sources are fundamental. New mechanisms for the efficient social collection of data, from the web as well as from established sources are the backbone of the PNS. These may harness diverse and innovative mechanisms for capturing data, sharing data, undertaking analysis for both static and real time scenarios across a wide range of potential application areas.

\item Ability to dynamically assemble and configure an ensemble of information sources needed to fulfil a high level goal. As key pillar of this process will the system’s social awareness that will allow it to understand which resources can best utilized under what circumstances, which users may be willing to contribute data to which causes, and to prioritize different requests according to social urgency. An integrated ontology accounting for information flows within socio-economic systems is a crucial enabler for such awareness.

\item Ability to infer the required information from the data in a distributed, bottom up way that refrains from storing any individually identifiable information. Social awareness will be used to derive priors and instantiate high level models that will support the reasoning. Techniques to verify that the privacy cannot be violated from aggregation of independent anonymous data are a further important factor.

\item Ability to build up and efficiently maintain a hierarchical, distributed and scalable knowledge base resulting from the population of the integrated ontology, capturing the current state and internal structure of global social systems as a basis for social awareness, and identifying as well the mechanisms for social adaptation. The knowledge base will combine information gathered and inferred by the PNS itself with knowledge returned by the LEV.
\end{enumerate}

In a nutshell, the PNS we envisage is a techno-social ecosystem that, when provided with a high-level analytical request, activates a global, distributed social knowledge discovery process to fulfil the request. An example of this process is how the PNS can support the computation and monitoring of new indices of social well-being, a novel compass long-awaited by decision-makers. It has often been pointed out that the gross national product (GDP) per capita is not a sufficient compass for politics to promote social well-being. Along this trend, the French presidency has recently established a high-level expert committee to identify better indices, involving Nobel prize winners such as Joseph Stiglitz, Amartya Sen, Kenneth Arrow, and Daniel Kahnemann \cite{StiglitzSen1,StiglitzSen2}. Their proposal is that, besides material living standards (income, consumption and wealth), it is important to consider many other factors, such as health, education, personal activities including work, political voice and governance, social connections and relationships, environment (present and future conditions), security, of an economic as well as a physical nature. It should be underlined that these factors are not only to be understood as concession to the people. They have also a significant impact on economic wealth creation, which depends not merely on material factors (such as suitable location, sufficient capital, well-functioning supply networks etc.), but also on human and social capital.

It is equally important to stress that finding a better index than GDP per capita is just one problem. Maximizing a one-dimensional index is another one. While it may promote an increase in average performance, it will necessarily produce winners and losers. The weakness of the concept of competitive optimization in one dimension is dramatically evidenced by the current struggle of several European countries to avoid bankruptcy.
An alternative concept, namely individually favorable optimization, has been proposed in \cite{HB2011}, particularly Sec. 5.2 (Towards indices of human well-being), Sec. 5.3 (The economics of happiness), Sec. 5.4 (New incentive systems), and Appendix A (Pluralistic indices promoting individual talents). Rather than optimizing a single index that is defined by a weighted average, it puts higher weights on individual strengths and less on individual weaknesses. This approach appreciates individual excellence and diversity, acknowledging their relevance for happiness, innovation, and systemic resilience. In other words, considering many different dimensions of social well-being and weighting them according to local and regional strengths allows for specialization according to local or regional excellence. Such a multi-dimensional approach supports not only competitiveness and complementarity, but also social well-being, and it is perfectly compatible with different local and regional cultures, while one-dimensional optimization of an index is not.
One of the scientific challenges in creating indices for social well-being will be to measure the relevant factors globally and in real-time with sufficient accuracy. (Delayed policy response may cause an unstable system dynamics.)

Measuring social well-being is even more difficult than measuring GDP. Currently, reliable official numbers for GDP are published only with a delay of many months. However, new ways of measuring GDP have recently been suggested. For example, it seems feasible to estimate GDP based on satellite pictures of global light emissions \cite{GDPestimation}. Such estimates are possible almost in real-time. Similarly, it has been shown that health-related indicators (such as the number of patients during flu pandemics) can be well estimated based on Google trends data \cite{GoogleFluTrends}.

Therefore, the vision of FuturICT is to make the different dimensions of social well-being globally measurable in real-time. This could be done by social sensing and mining from disparate accessible sources, as described in this paper. Recent attempts to measure happiness and its variation in space, time and across social communities point the way for this \cite{DoddsHappinessIndex,MichaelMacyScience}. At present, the greatest difficulty in measuring human capital (such as education) or social capital (such as cooperativeness and solidarity) is the calibration and validation of corresponding indices. It is envisaged to do this by comparison with (online) surveys and (web) experiments.
FuturICT’s “Planetary Nervous System” (PNS) will provide the methods and tools to measure human activities and socially relevant variables in real-time, on a global scale, and in a privacy-respecting way. By extending measurements to social and economic domains, FuturICT will complement and go beyond similar projects focused on environmental and climate-oriented measurements, e.g. Planetary Skin \cite{PlanetarySkin} and Digital Earth \cite{DigitalEarth}. But the ambition of the Planetary Nervous System goes beyond measurement. It also intends to create collective awareness of the impact of human decisions and actions, particularly on the social fabric on which our society is built (the “social footprint”) \cite{N1}.

The Planetary Nervous System serves to detect possible opportunities and threats, in order avoid mistakes that we may regret later on. This requires a certain ability to anticipate (in a probabilistic way) possible courses of events (even if this anticipation is unreliable and only useful for short time periods, such as our weather forecast). While our own consciousness performs such anticipation by “mental simulation”, FuturICT’s “Living Earth Simulator” (LES) will do the equivalent task for complex techno-socio-economic-environmental systems, building on data provided by the Planetary Nervous System. These simulations will be based on models of our society and economy and other relevant activities in our world. They may be imagined to work like a “policy wind tunnel” to explore possible impacts of human decisions and actions, both positive and negative ones.
The Planetary Nervous System and Living Earth Simulator will be connected to the ‘Global Participatory Platform’ (GPP), which will make FuturICT’s new methods and tools available for everyone (with mechanisms in place to foster responsible use). This will enable people to look at all interesting issues from many angles and to use the power of crowd sourcing and the wisdom of crowds.

While FuturICT’s new concepts for the measurement, simulation and interactive exploration of the impact of human decisions and actions will promote collective awareness, the Global Participatory Platform will create new opportunities for social, economic, and political participation. Altogether (and with suitable rating and reputation mechanisms in place), this will promote better decisions and responsible actions. In particular, measuring the value of human and social capital and quantifying the “social footprint” will help one to protect the social fabric on which our society is built, in a similar way as the measurement of the “environmental footprint” has empowered people and institutions to better protect our environment.

\subsection*{Acknowledgements}
The authors are indebted with many people for inspiring discussions, contributions, comments and criticisms, and remarkably: Andrzej Nowak, John Shawe-Taylor, Andreas Krause, Frank van Harmelen, Carole Goble, Wendy Hall, Sandra Hirche, Nicola Guarino, Giuseppe Attardi, Salvatore Orlando, Marco Conti, Chiara Renso, Piero Fraternali, Themis Palpanas, Stefano Leonardi. The publication of this work was partially supported by the European Union's Seventh Framework Programme (FP7/2007-2013) under grant agreement no.284709, a Coordination and Support Action in the Information and Communication Technologies activity area (`FuturICT' FET Flagship Pilot Project)."

\end{document}